\documentclass[journal=jacsat,manuscript=article]{achemso}

\usepackage{chemformula} 
\usepackage[T1]{fontenc} 

\author{Manuel Blanco}
\affiliation[USC]
{Grupo de investigaci\'on Photonics4Life, \'Area de \'Optica, Departamento de F\'isica Aplicada, Universidade de Santiago de Compostela, Santiago de Compostela, Spain}
\author{Ferran Cambronero}
\affiliation[USC]
{Grupo de investigaci\'on Photonics4Life, \'Area de \'Optica, Departamento de F\'isica Aplicada, Universidade de Santiago de Compostela, Santiago de Compostela, Spain}

\author{Mar\'ia Teresa Flores-Arias}
\affiliation[USC]
{Grupo de investigaci\'on Photonics4Life, \'Area de \'Optica, Departamento de F\'isica Aplicada, Universidade de Santiago de Compostela, Santiago de Compostela, Spain}
\author{Enrique Conejero Jarque}
\affiliation[USAL]
{Grupo de Investigaci\'on en Aplicaciones del L\'aser y Fot\'onica, Departamento de F\'isica Aplicada, University of Salamanca, E-37008, Salamanca, Spain}
\author{Luis Plaja}
\affiliation[USAL]
{Grupo de Investigaci\'on en Aplicaciones del L\'aser y Fot\'onica, Departamento de F\'isica Aplicada, University of Salamanca, E-37008, Salamanca, Spain}
\author{Carlos Hern\'andez-Garc\'ia}
\affiliation[USAL]
{Grupo de Investigaci\'on en Aplicaciones del L\'aser y Fot\'onica, Departamento de F\'isica Aplicada, University of Salamanca, E-37008, Salamanca, Spain}
\email{carloshergar@usal.es}

\title[ultrafast magnetism]
  {Ultraintense femtosecond magnetic nanoprobes induced by azimuthally polarized laser beams}


\begin{document}



 

%
%
%
%
%

\begin{abstract}
We report a novel scheme to generate laser-induced, ultrafast, intense (Tesla scale), spatially isolated, magnetic fields. Three-dimensional particle-in-cell simulations show that a femtosecond azimuthally-polarized infrared vector beam, aimed to a conducting circular aperture, produces an intense axially polarized tip-shaped femtosecond magnetic field, extending over micrometer distances and being isolated from the electric field.  Our results are backed-up by an analytic model, demonstrating the underlying physics and guiding for optimal parameters.
In particular, we find the conditions under which the magnetic nanoprobe is substantially enhanced, reaching 4 T when driven by a $10^{11}$ W/cm$^2$ laser field, which reflects a selective enhancement by a factor of $\sim$6.
Our scheme offers a promising tool to control, probe and tailor magnetic nanodomains in femtosecond timescales through pure magnetic interaction by using structured laser beams.
\end{abstract}

The development of new techniques to accurately control the spatial and temporal properties of materials is one of the most exciting challenges in current nanotechnology, and a required step towards the next generation of electronic devices, magnetic storage, or photovoltaic materials, among others. 
In this context, current laser technology offers unique tools in the form of ultrashort pulses --as short as few femtoseconds (1 fs=$10
^{-15}$ s), or even attoseconds ($10
^{-18}$ s) 
\cite{Krausz2009}--, that can be focused down to the nanometer scale \cite{Popmintchev2012}, and that can be structured in their angular momentum properties (polarization \cite{Huang2018} and orbital angular momentum \cite{Hernandez-Garcia2013}). 
Ultrashort laser sources have already shown great potential for applications in ultrafast magnetism. Since the pioneering work on ultrafast laser-induced demagnetization by Beaurepaire et al. \cite{Beaurepaire1996}, fs laser pulses have been widely used in theoretical and experimental studies of femtomagnetism 
\cite{Kimel2005,Koopmans2009,Bigot2009, Boeglin2010,Rudolf2012, Turgut2013, Chen2017,Tengdin2018}. The magnetic effects in such short time-scales appear not only due to thermal relaxation processes, but also to coherent and almost instantaneous interactions between photons and spins. Such dynamics can only be tested using ultrashort laser pulses. 

A remarkable feature of the interaction of laser pulses with magnetic materials is its strong dependence with the polarization \cite{Kimel2005,Stanciu2007, Fan2015}. 
During the recent years the development of vector beams --laser sources with structured polarization-- has opened new avenues for magnetic laser-matter interaction. In particular, azimuthally polarized vector beams possess enormous potential to control pure magnetic interactions \cite{Veysi2015,Veysi2016}. Previous works have demonstrated the feasibility of azimuthally polarized laser beams to induce static magnetic fields --which can be enhanced towards few mili-Tesla through the use of nanoantennas \cite{Guclu2016}--, with applications in nanoscale photoinduced force microscopy \cite{Zeng2018}.
With the development of ultrashort laser vector beams \cite{Hernandez-Garcia2017} we envision the opportunity to perform pure magnetic ultrafast interactions, including ultrafast demagnetization or spin precession. However, such interactions require the use of intense magnetic fields.  Although in the last years Tesla-scale magnetic fields have been reported at the nanosecond and picosecond timescales, --through laser-plasma interaction \cite{Sandhu2002}, or benefiting from transient thermoelectric currents in metals \cite{Tsiatmas2013, Vienne2015}--, to the best of our knowledge, there is no evidence of the generation of such intense, isolated, magnetic fields in the fs timescale.

In this work we propose a novel technique to generate intense, femtosecond longitudinal magnetic fields, spatially isolated from the electric field, using vector laser beams. 
We perform a complete analysis of the spatio-temporal properties of such magnetic pulses through the use of three-dimensional (3D) particle-in-cell (PIC) simulations and the development of an analytic model. In particular, we show that the longitudinally polarized magnetic field finds a maximum amplitude at the aperture axis, being isolated from the electric field, and creating an ultrafast magnetic nanoprobe with a tip-like shape that extends over micrometer distances. 


\begin{figure}
  \begin{center}

   \includegraphics[width=12.8cm]{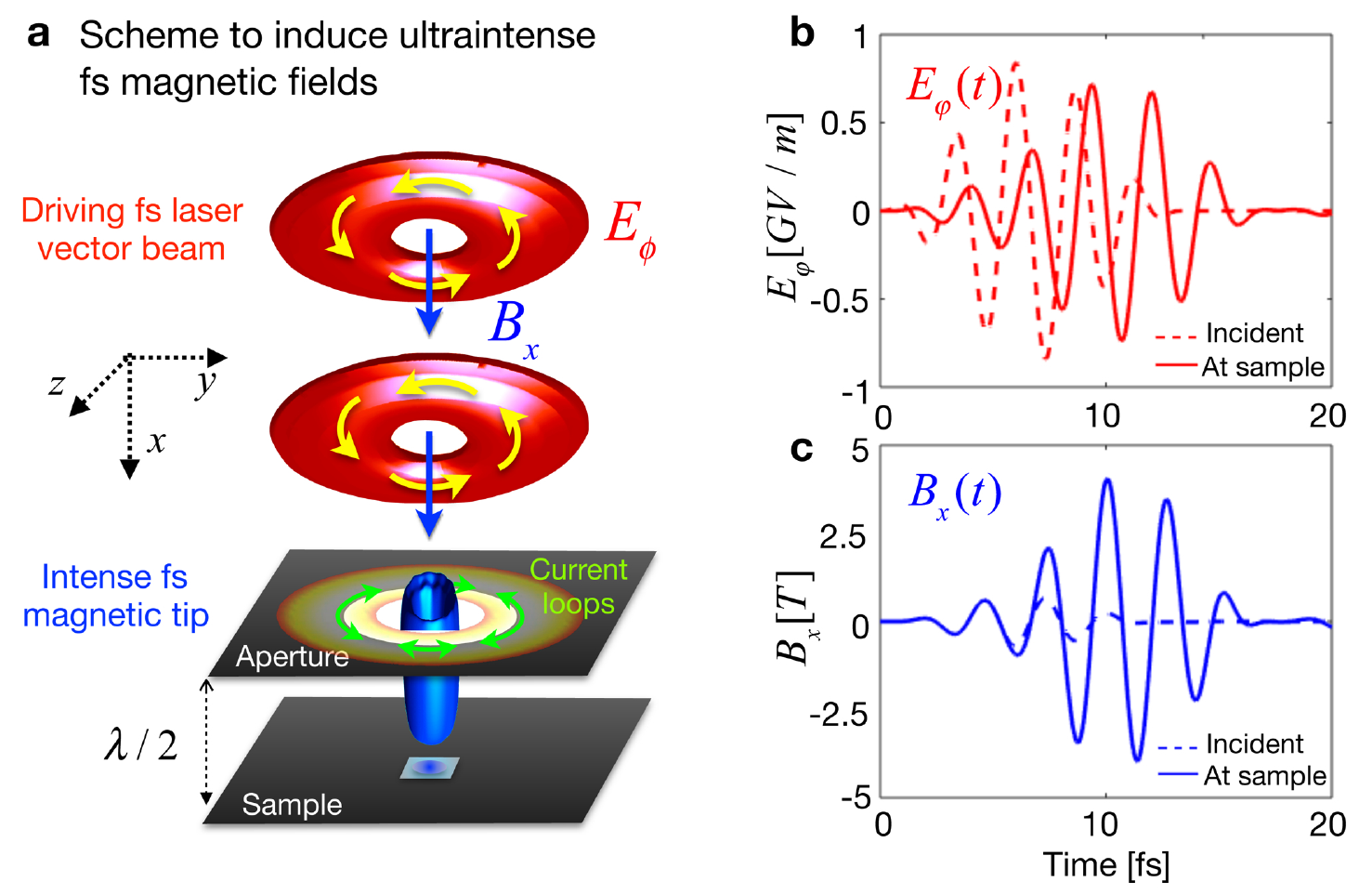}
  \caption{(a) Scheme to generate tip-shaped ultraintense ultrafast longitudinal $B$ fields. An azimuthally polarized laser beam (red) is aimed to a metallic circular aperture. The current loops (green arrows) induced at the edge of the aperture enhance the longitudinal $B$ field (blue) located at the singularity of the vector beam, resulting in a tip-shaped ultraintense ultrafast $B$ field of micrometer scale. The dashed lines In panels (b) and (c) show the temporal profile of the incident azimuthal electric field ($E_\varphi$) at $\rho=\rho_0$, and the incident longitudinal magnetic field ($B_x$) on-axis ($\rho= 0$), respectively, detected at the focal plane ($x=0$), and obtained through 3D PIC simulations. The solid lines represent the $E_\varphi$ and $B_x$ fields detected at a sample plane located at $x=\lambda/2$ when placing the circular aperture. When comparing the incident and apertured fields, whereas $E_\varphi$ remains similar, the apertured $B_x$ field is enhanced by a factor of $\sim 6$ (reaching 4 T). Note that the $E_\varphi$ fields presented in (b) are detected at the spatial positions of maximum intensity for each case --$\rho=\rho_0$ for the incident field and $\rho=0.24$ $\mu$m for the apertured one--, which is responsible for the temporal delay observed.}
  \label{fig:fig1}
    \end{center}

\end{figure}


The electric field ($E$) of an azimuthally polarized beam can be described in the paraxial approximation as a Laguerre-Gaussian mode (LG$_{1,0}$) without the azimuthal phase \cite{Zhan2009}. Solving the Maxwell equations, one can obtain the corresponding magnetic field ($B$) \cite{Veysi2015}. Thus, considering an azimuthally polarized laser beam propagating along the $x$ direction, the electromagnetic field obtained at the focal plane ($x=0$) and expressed in cylindrical coordinates $(\rho,\varphi, x)$ reads as
\begin{eqnarray}
\vec{E}(\vec{r}) &=& E_0 \frac{\rho}{w_0} e^{-\frac{\rho^2}{w_0^2}} \vec{e}_{\varphi} \label{eq:efield} \\
\vec{B}(\vec{r}) &=& -\frac{E_0}{c} \frac{\rho}{w_0} e^{-\frac{\rho^2}{w_0^2}} \vec{e}_{\rho} - 2 \frac{E_0}{c} \frac{\lambda}{2 \pi w_0} \left[1 - \frac{\rho^2}{w_0^2}\right] e^{-\frac{\rho^2}{w_0^2}} e^{i \pi / 2} \vec{e}_x
\label{eq:bfield}
\end{eqnarray}
where $E_0$ is the $E$ field amplitude and $\vec{e}_\varphi$, $\vec{e}_\rho$ and $\vec{e}_x$ are the unitary vectors in the cylindrical coordinate system. The beam waist is $w_0$; $\lambda$ is the wavelength; and $c$ is the speed of light. 

Figure \ref{fig:fig1}a shows the interaction scheme considered: the azimuthally polarized beam is aimed perpendicularly to a metallic circular aperture.
We perform numerical simulations using the OSIRIS 3D PIC code \cite{Fonseca2003,Fonseca2008,Fonseca2013} to analyze the spatio-temporal behavior of the electromagnetic field with and without the metallic aperture.  
We have considered as an incident laser pulse the azimuthally-polarized beam described by Eqs. (\ref{eq:efield}) and (\ref{eq:bfield}), with waist $w_0=2.5$ $\mu m$, a central wavelength of $\lambda=0.8$ $\mu m$, and $E$ field amplitude of 0.86 GV/m (peak intensity of $10^{11}$ W/cm$^2$) at the radius of maximum intensity ($\rho_0=w_0/\sqrt{2}$). The temporal envelope is modeled as a sin$^2$ function of 4.8 fs full width at half maximum in intensity. In order to implement the proposed scheme for the amplification of the $B$ field, we have considered thin gold circular apertures (density of $5.9\times 10^{22}$ cm$^{-3}$, and 50 nm in thickness) with different radii, $\rho_A$, placed at the beam focus position ($x=0$). Note that the maximum peak intensity of the laser pulse has been chosen to be below the damage threshold of the material of the aperture.

The central outcome of our work is shown in Figs. \ref{fig:fig1}b and \ref{fig:fig1}c, where we present the $E_\varphi$ field at the radius of maximum intensity and the $B_x$ at $\rho= 0$ of the incident (dashed lines) and apertured (solid lines) beams. Whereas the incident fields are shown at the focal plane $x=0$, the apertured fields are shown at a plane located at $x=\lambda/2$. As a main result, we observe that the incident beam naturally induces an ultrafast $B_x$ field that oscillates at 0.37 PHz, with a maximum amplitude of 0.7 T (dashed blue line in Fig. \ref{fig:fig1}c). Interestingly enough, when placing a gold circular aperture with radius equal to that of maximum intensity of the beam, $\rho_A=\rho_0=1.78$ $\mu$m, the $B_x$ amplitude is enhanced by a factor of $\sim6$ with respect to the non-apertured case, reaching a peak value of 4 T.

 \begin{figure}
   \begin{center}

    \includegraphics[width=16.5cm]{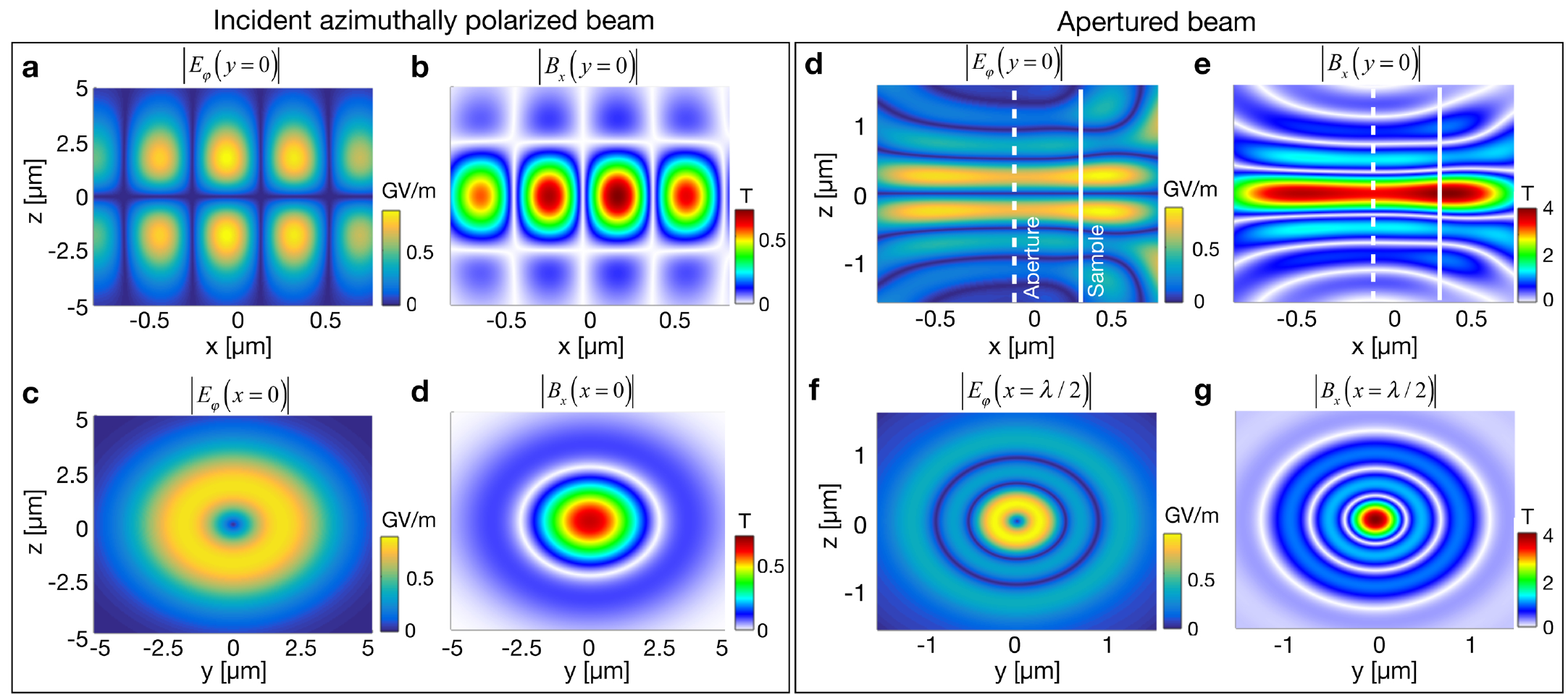}
   \caption{Spatial distributions of the $E_\varphi$ and $B_x$ fields carried by an azimuthally polarized laser pulse beam (left) without and (right) with a gold circular aperture of radius $\rho_A=\rho_0=1.78$ $\mu m$, placed at $x=0$. (a,d) and (b,e)  show the longitudinal spatial profiles of $E_\varphi$ and $B_x$ at $y=0$, respectively. (c) and (d) show the transverse ones at the focal plane $x=0$ for the incident beam, whereas (f) and (g) show the corresponding transverse profiles of the apertured $E_\varphi$ and $B_x$ fields at the sample plane $x=\lambda/2$, respectively. The aperture (white dashed) and sample (white solid) positions are indicated in (d) and (e). These spatial profiles are obtained through 3D PIC numerical simulations.}
     \label{fig:fig2}
     \end{center}

 \end{figure}

In Fig. \ref{fig:fig2} we show the incident (left) and apertured (right) fields at the longitudinal (top) and transverse (bottom) planes obtained from our 3D PIC simulations. The spatial distributions are presented at the temporal instant when each field is maximum. 
When placing a gold aperture of radius $\rho_A=\rho_0$, the $E_\varphi$ and $B_x$ fields are strongly focused, but, remarkably, both their longitudinal and transverse distributions remain separated (right side of Fig. \ref{fig:fig2}). Note that the radial magnetic field, $B_\rho$, is also spatially separated from $B_x$ (see Supporting Information).

\begin{figure}
  \begin{center}
   \includegraphics[width=10.5 cm]{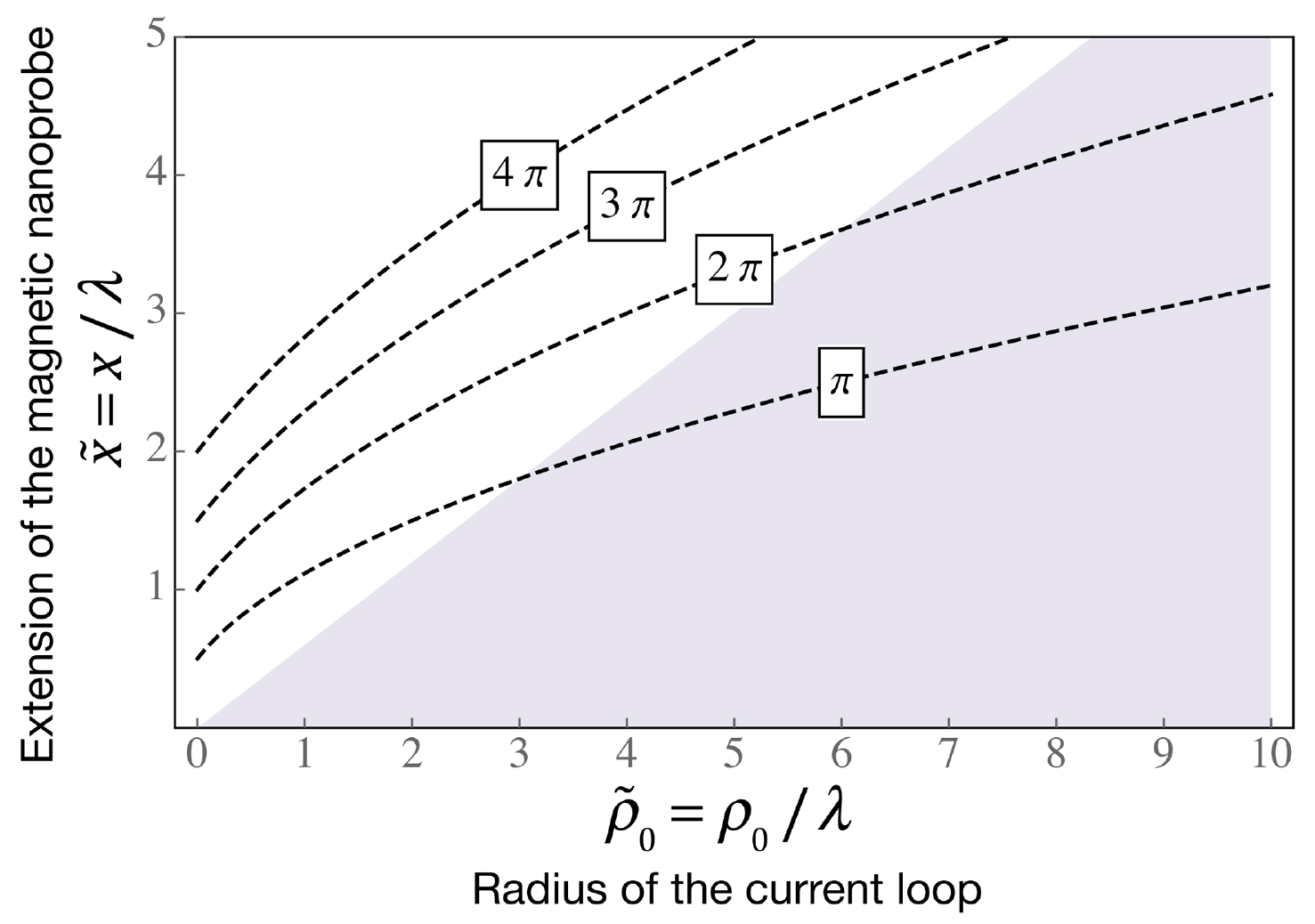}
  \caption{Extension and phase of the  magnetic nanoprobe as a function of the radius of the current loop induced in the aperture, obtained with the analytic model described by Eqs. (\ref{eq:modB}) and (\ref{eq:phaseB}). The grey area shows the extension of the magnetic nanoprobe ($\ell \simeq 0.6 ~ \rho_0$) defined as the separation from the plate for which the $B_x$ field energy drops by a factor of 2. The contour dashed lines show the spatial phase shift of the $B_x$ field.}
  \label{fig:fig3}
  \end{center}
\end{figure}

Another remarkable feature of the $B$ field profile after the aperture is its narrow width in comparison with its longitudinal extension, forming a tip-shaped probe, as depicted in Fig. \ref{fig:fig2}e. We have developed a simple model to characterize the spatial profile of this magnetic nanoprobe by analyzing the field induced by the alternate electric azimuthally polarized field in the circular plate. The model simplifies the picture assuming a thin metallic plate and an induced current with a narrow ring-shaped form (see Supporting Information). The current loop induces a $B_x$ field with maximum amplitude at the beam propagation axis, that can be described by $B_{x,0}= |B_{x,0}| e^{i \varphi_{0}}$,  where
\begin{eqnarray}
|B_{x,0}(\tilde{\rho_0};\tilde{x})| & \propto & \frac{1}{\lambda} \frac{\tilde{\rho}_0^2}{\tilde{\rho}_0^2+ \tilde{x}^2} \left( 4 \pi^2 - \frac{1}{\tilde{\rho}_0^2+ \tilde{x}^2}\right)^{1/2} \label{eq:modB} \\
\varphi_0(\tilde{\rho_0};\tilde{x},t) &=& 2 \pi \sqrt{\tilde{\rho_0}^2+\tilde{x}^2}-\arctan{\left ( 2 \pi \sqrt{\tilde{\rho_0}^2+\tilde{x}^2} \right )} - \omega t  \label{eq:phaseB}
\end{eqnarray}
being $\tilde{\rho_0}=\rho_0/\lambda$ and $\tilde{x}=x/\lambda$ the normalized radial and longitudinal coordinates. We define the probe length, $\ell$, as the distance from the plate for which the $B$ field energy drops by a factor of two. Imposing this condition to the squared modulus of Eq. (\ref{eq:modB}), we find a simple scaling of the probe length with the radius of the current loop, $\ell \simeq 0.6 ~ \rho_0$. Thus, the probe typically extends over distances exceeding a few $\lambda$, allowing for almost pure magnetic interactions close to the axis of a target plane located at micrometer distances from the aperture. From the practical point of view, it is desirable  a nearly uniform distribution of the $B_x$ field over the thickness of a potential target. Therefore, it becomes necessary that the phase of the field amplitude of the magnetic probe does not change rapidly along the target thickness dimension. We show in Fig. \ref{fig:fig3} the probe extension as a function of the current loop radius --indicated by the grey area-- and the spatial phase distribution --indicated by the contour dashed lines--. It is worth to note that for a loop radius of $3\lambda$, the field energy is confined to a region extending over $\simeq1.8\lambda$, with a phase variation $<\pi$.

 \begin{figure}
   \begin{center}

    \includegraphics[width=11cm]{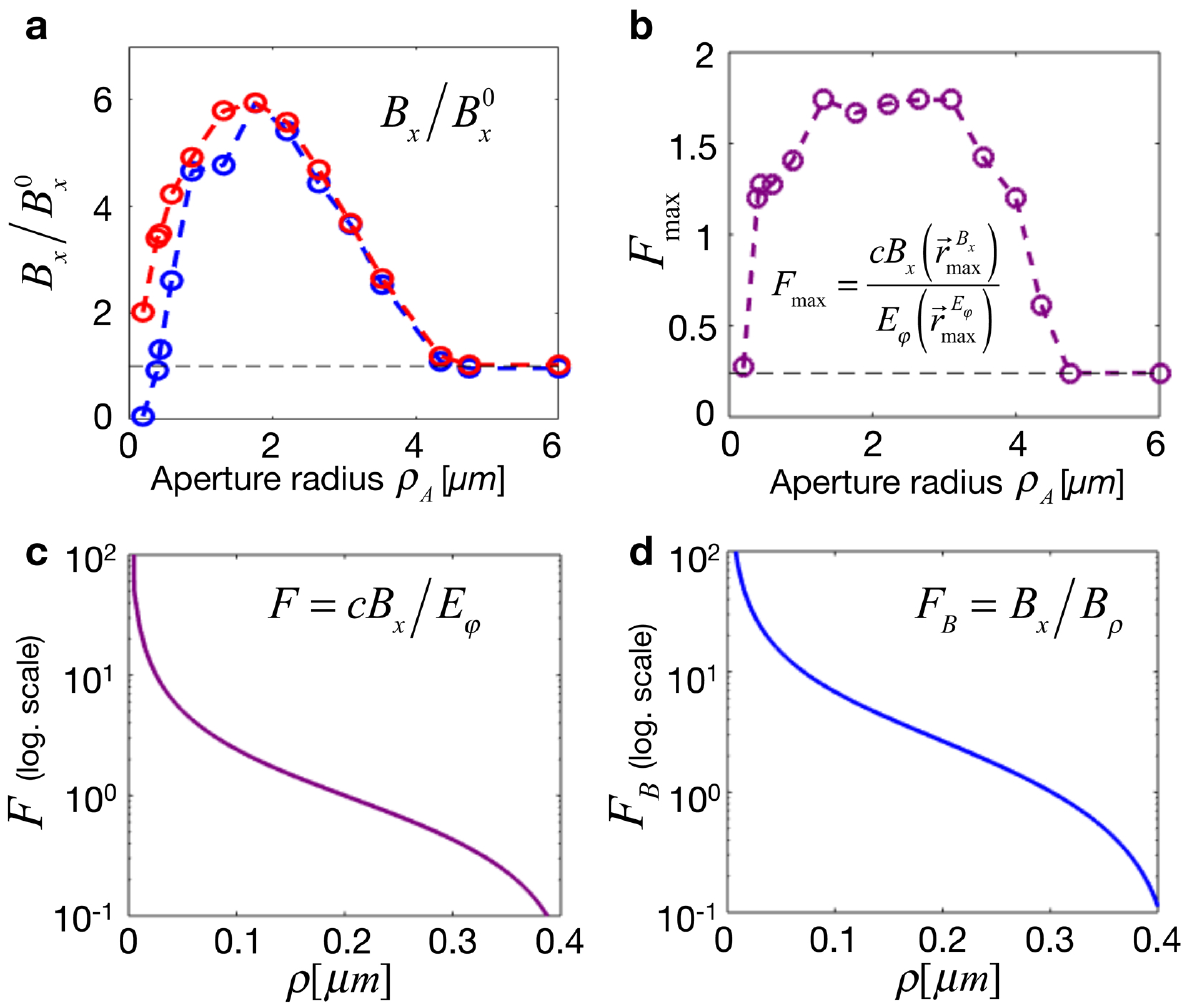}
   \caption{(a) Amplification of the $B_x$ field for different aperture sizes. The ratio $B_x/B_x^0$ at $\rho=0$ is presented for different aperture radii, where $B_x$ is measured at the sample ($x=\lambda/2$, blue) and at the spatial position where it is maximum (red), and $B_x^0$ is the incident field without the aperture. (b) Contrast between the maximum $B_x$ and $E_\varphi$ fields, i.e. $F_{max}=cB_x(\vec{r}^{B_x}_{max})/E_\varphi(\vec{r}^{E_\varphi}_{max})$, where $\vec{r}^{B_x}_{max}$ ($\vec{r}^{E_\varphi}_{max}$) is the position where $B_x$ ($E_\varphi$) is maximum. The black dashed line indicate the value for the non-apertured case. The maximum amplification and contrast $F_{max}$ take place for an aperture radius $\rho_A=\rho_0$. (c) Radial profile of the contrast $F(\rho)=B_x(\rho)/E_\varphi(\rho)$ and (d) that of $F_B(\rho)=B_x(\rho)/B_\rho(\rho)$, which allow us to define the spatial region in which the $B_x$ field is isolated from the $E_\varphi$ and $B_\rho$ fields.}
   \label{fig:fig4}
    \end{center}

 \end{figure}

Our analytic model indicates that the electronic current induced at the edge of the aperture --and thus the induced $B_x$ field-- reaches its maximum when the aperture radius matches that of the ring of maximum intensity of the vector beam, i.e., $\rho_A=\rho_0$ (see Supporting Information). This is confirmed by the results from 3D PIC simulations shown in Fig. \ref{fig:fig4}(a), where
we present the ratio $B_x/B_x^0$ --being $B_x^0$ is the incident field without the aperture-- at $\rho=0$ for different aperture radii, measured at the sample plane ($x=\lambda/2$, blue) and at the spatial position of maximum $B_x$ (red). 
In addition, the aperture radius introduces a proportional temporal delay to the induced $B_x$ field, as it can be observed in Fig. \ref{fig:fig1}(c). Movies in the Supporting Information show the temporal evolution of the build-up of the $E_\varphi$ and $B_x$ fields for different aperture radii.

At these high intensities it is interesting not only to increase the $B$ field amplitude, but also the contrast between the $B$ and $E$ fields, which we define as $F=c|B_x|/|E_\varphi|$. In the 3D PIC simulations presented in Figs. \ref{fig:fig1} and \ref{fig:fig2}, we observe that, whereas the $B_x$ field is amplified when placing the aperture, the $E_\varphi$ field amplitude remains similar, resulting in an increased contrast $F$. This is a remarkable aspect of our scheme, as it allows to increase the $B$ field avoiding high $E$ fields that could reach the damage threshold of the sample. Our simulations also show that the contrast between the maximum $B_x$ and $E_\varphi$ fields --$F_{max}$, shown in Fig. \ref{fig:fig4}(b)-- is maximized for an aperture of radius $\rho_A=\rho_0$.

Finally, in order to define the spatial region where the $B_x$ field can be considered as isolated, we present in Fig. \ref{fig:fig4}(c) the radial profile of the contrast between the $B_x$ and $E_\varphi$ fields, $F(\rho)=cB_x(\rho)/E_\varphi(\rho)$, and in Fig. \ref{fig:fig4}(d) that between the $B_x$ and $B_\rho$ fields, $F_B(\rho)=B_x(\rho)/B_\rho(\rho)$. Note that these curves are very similar for all aperture radii. From  Figs. \ref{fig:fig4}(c) and \ref{fig:fig4}(d) one can define a spatial region in which the $E_\varphi$ and/or the $B_\rho$ field can be neglected with respect to the $B_x$ field. Note that the size of this region could be modified through the use of different beam waists ($w_0$), keeping in mind that the optimal aperture for amplifying the $B_x$ field with respect to the $E_\varphi$ satisfies $\rho_A=\rho_0$.

In conclusion, we present a scheme to generate ultraintense, ultrafast, isolated longitudinal magnetic fields from azimuthally polarized laser beams. The use of a thin circular aperture allows us to amplify the longitudinal magnetic field by a factor of $\sim$6, reaching 4 T for an incident laser vector beam of $10^{11}$ W/cm$^2$, --being thus comparable to the magnetic field amplitudes of the large bending magnets used in synchrotron facilities--. We also have shown that the circular aperture enhances the contrast between the $B$ and $E$ fields, allowing for the isolation of the $B$ field in regions of several $nm^2$. The $B$ field is found as a tip-shaped probe whose length scales linearly with the aperture radius, allowing to implement pure magnetic interaction with a material target placed at micrometer distances from the plate. Thus, our scheme offers the possibility to tailor the spatial and temporal properties of the $B$ field through proper modifications of the incident laser beam and metallic aperture parameters.

Our findings present potential applications in different fields. On one hand, 
instead of controlling material magnetic properties with the $E$ field of a fs laser pulse, we envision the possibility of controlling the magnetic properties of materials with a fs $B$ field. For example, a fast magnetization reversal --which is of paramount importance for the realization of high-rate magnetic recording-- could be triggered at the fs scale with $B$ fields of hundreds of T. One could also advance the possibility of generating attosecond isolated $B$ fields if shorter wavelength ultrashort vector beams \cite{Hernandez-Garcia2017} were used.
On the other hand, ultraintense, ultrafast $B$ nanoprobes can be applied to improve the capability to control the acceleration of charged particles in laser-plasma interaction schemes \cite{Vieira2011, Rassou2015, Korneev2017} 
, to optimize the generation of coherent extreme-ultraviolet radiation through high harmonic generation \cite{Milosevic1999} or to enhance the betatron oscillation amplitude in the generation of x-rays \cite{Zhang2016}.

\begin{acknowledgement}

The authors thank Luis L\'opez, Eduardo Mart\'inez and Victor J. Raposo for fruitful discussions. 
The authors thank support from Ministerio de Econom\'ia y Competitividad (FIS2015-71933-REDT) and Xunta de Galicia/FEDER (ED431B 2017/64). C.H.-G. acknowledges support from a 2017 Leonardo Grant for Researchers and Cultural Creators, BBVA Foundation. C.H.-G., E.C.J. and L.P. acknowledge support from Junta de Castilla y Le\'on (SA046U16) and Ministerio de Econom\'ia y Competitividad (FIS2016-75652-P). M. B. is funded by FPU grant program of MECD. The authors acknowledge the OSIRIS Consortium, consisting of UCLA and IST (Lisbon, Portugal) for the use of OSIRIS, for providing access to the OSIRIS framework.

\end{acknowledgement}

\begin{suppinfo}

\begin{itemize}
  \item SupportingInformation.pdf: it includes  a detailed discussion of: (I) the parameters of the 3D PIC simulations, and the spatial and temporal profiles of the radial component of the $B$ field; (II) the temporal evolution of the spatial profile of the $E$ and $B$ fields propagating through different apertures; and (III) the complete derivation of the analytic model that describes the sparial properties of the magnetic nanoprobe.
  \item EphiXZapertured.avi: temporal evolution of the spatial distribution of the $E_\varphi$ field for four different aperture radii.
\item BxXZapertured.avi: temporal evolution of the spatial distribution of the $B_x$ field for four different aperture radii.
\end{itemize}

\end{suppinfo}

\bibliography{NanoMagnetic}

\end{document}